# Traffic Confirmation Attacks Despite Noise

Jamie Hayes
University College London
j.hayes@cs.ucl.ac.uk

***Abstract*—We propose a traffic confirmation attack on low-latency anonymous communication protocols based on computing robust real-time binary hashes of network traffic flows. Firstly, we adapt the Coskun-Memon Algorithm to construct hashes that can withstand network impairments to allow fast matching of network flows. The resulting attack has a low startup cost and achieves a true positive match rate of 80% when matching one flow out of 9000 with less than 2% false positives, showing traffic confirmation attacks can be highly accurate even when only part of the network traffic flow is seen. Secondly, we attack probabilistic padding schemes achieving a match rate of over 90% from 9000 network traffic flows, showing advanced padding techniques are still vulnerable to traffic confirmation attacks.**

## I. INTRODUCTION

Internet communication traffic is commonly encrypted to hide its content using TLS. However, TLS encrypted traffic is vulnerable to traffic analysis since it does not hide packet metadata, such as the time a packet was sent or received or the size of the packet.
Low-latency mix networks attempt to provide anonymous communication by obscuring the flow of traffic through the network. Intermediate mixing of messages removes sender-receiver linkability. An Onion Routing (OR) network obscures sender-receiver linkability in a similar fashion but security is obtained through route unpredictability; mixing of messages is not typically required. A client that browses to a website through an OR network [8] will have their traffic relayed before it reaches its destination. Each relay only knows where to send the traffic next, so under the assumption that relays do not collude with one another, an adversary observing traffic at a relay will not be able to infer both the origin and destination of the clients communications. Though low-latency mix networks and OR networks differ in how they operate, both are susceptible to timing attacks because communication patterns are not intentionally delayed for long periods.

We consider a passive traffic confirmation attack, where an adversary eavesdrops on two ends of the network and aims to link the sender and receiver of a communication over the OR network. Research has shown that if an adversary can view traffic at both the entry and exit relay, traffic confirmation attacks are possible [9], [11], [14], [15], [19], [21]. However doubt has been cast upon the efficiency of traffic analysis over OR networks since attack accuracy is expected to suffer due to background noise and network conditions. The base rate fallacy, the cost of exfiltration, processing and storage of data are expected to be contributing factors that negatively affect attack accuracy [15]. We present a new traffic confirmation attack which mitigates some of these fears, where an adversary can learn a short binary hash representation of a network flow in real-time and then compare it against a library of hashes recorded at another location in the network. Because an adversary only needs to store a short hash per network traffic flow the attack requires a low storage cost, the hashes are easy to exfiltrate from network traffic flows and provide a framework for computationally low cost matching and local evaluation. Our attack succeeds even when the adversary can only observe a fraction of the total transmission or when the network uses probabilistic padding schemes.

## II. BACKGROUND & MOTIVATION

Traffic confirmation attacks in low-latency mix networks are an active area of research [3], [6], [12] and fall in to two categories, passive and active attacks. In a passive attack an adversary aims to link a sender and receiver of a communication by inferring statistical properties of observed network traffic flows. In an active attack, commonly referred to as *tagging* or *watermarking*, the adversary modifies network traffic flows introducing patterns that be can observed at another location in the network, allowing the adversary to de-anonymize clients using the mix network [24]. Active attacks can be detected [16] or at least rendered no more powerful than passive attacks under padding schemes since padding may remove the signature imposed by the adversary on the flow.

Simple passive attacks such as counting the number of packets in network flows have been shown to yield strong results [19]. Levine et al. [11] shows that by dividing flows in to time windows and counting the packets within the windows a more powerful attack can be performed. Packet monitoring at scale becomes challenging due to adversary bandwidth restraints; Chakravarty et al. [5] showed that large scale attacks using less fine grained information is possible by monitoring network flow statistics as captured by servers such as Cisco's NetFlow. Note that the possibility of large scale traffic confirmation attacks is not merely an academic concern. As of October 2015 all ISPs in Australia must maintain NetFlow-like data for a minimum of two years [1]. One may wonder, if the adversary is a nation state or ISP with plenty of available bandwidth do they need a method for converting network traffic to short hashes? It has been shown that data exfiltration is expensive even for an adversary with a lot of bandwidth to utilize [10], [18]. Data is commonly compressed before exfiltration; a method for converting traffic signals to short hashes while preserving the contained information would greatly improve the speed and amount of data that can be exfiltrated and subsequently matched.

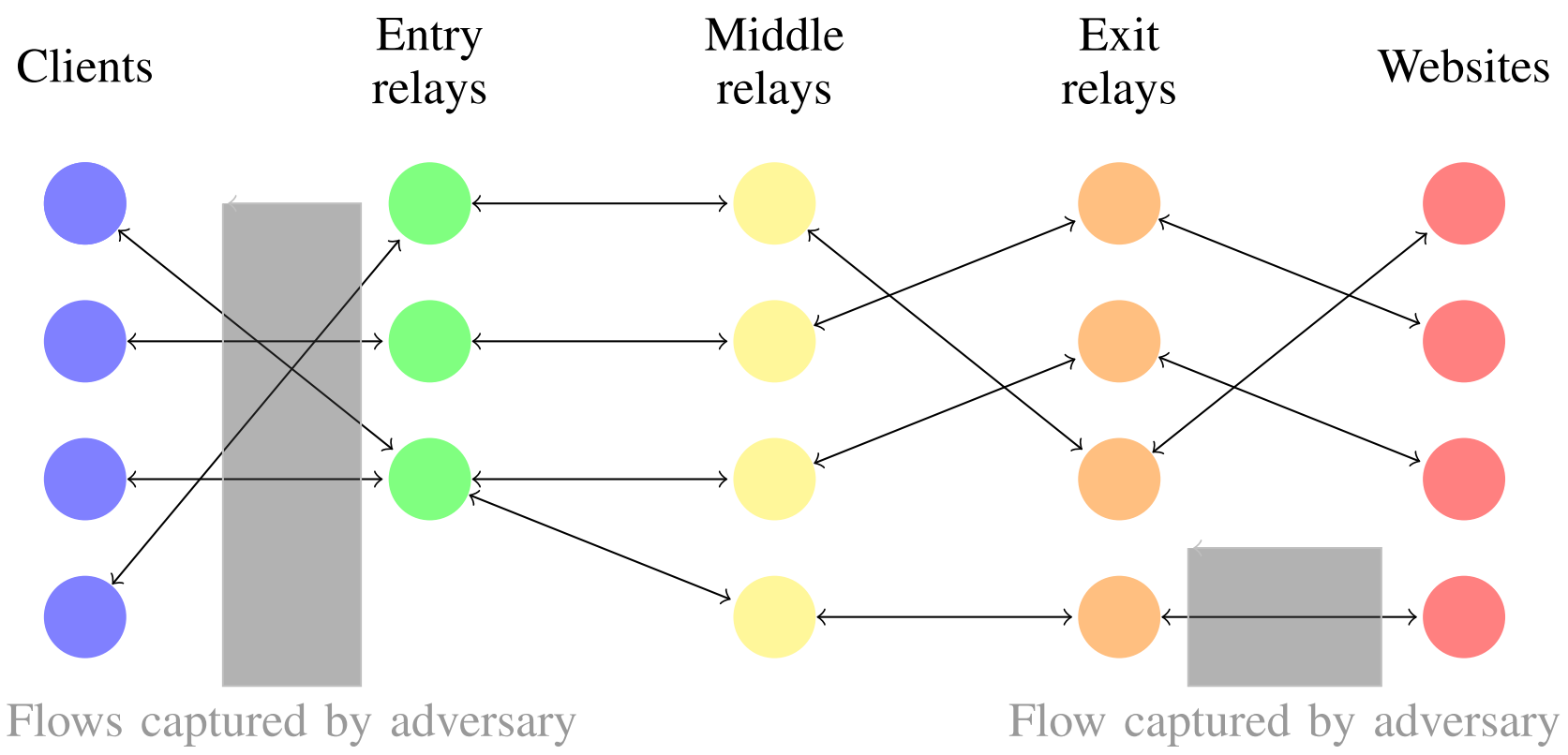

**Fig. 1:** Architecture of threat model.

Deterministic padding in OR networks while offering guarantees of security, are usually expensive in terms of latency and bandwidth overheads [22]. If constant rate padding is applied with the requirement of no latency overheads, packets must be injected at a rate less than or equal to the smallest inter-arrival time between packets. Since short-lived website connections are bursty in nature, adding constant rate cover traffic is likely to incur large bandwidth overheads. Probabilistic techniques such as Adaptive Padding (AP) [20] aim to protect anonymity by introducing traffic in to statistically unlikely delays between packets in a flow. This limits the amount of extra bandwidth required and does not incur any latency costs as packets are forwarded as soon as they are received. AP uses previously computed histograms of inter-arrival packet times from website loads to determine when a packet should be injected. The AP algorithm consists of two modes, gap and burst. When a network traffic flow has a natural delay in packets AP enters gap mode and increases the probability of injecting a dummy packet. This destroys natural fingerprints created by the gaps in flows. In burst mode, AP recognizes that traffic is flowing at a high rate, and so reduces the probability that a dummy packet will be injected.

We apply our attack to the most popular low-latency OR network, Tor [8]. Currently packets are sent through the Tor network using TCP which guarantees reliable transmission. However TCP flow control has been identified as a major cause of latency in Tor; there has been suggestions to incorporate the User Datagram Protocol (UDP) which would reduce queue lengths on relays and allow for better congestion management [13], [17]. If the reduction of latency in Tor or the use of VoIP and similar protocols is desired, it may be prudent to allow UDP over Tor. In this case packets may be dropped; motivating our study of attack tolerance when packets are not reliably transmitted over Tor. We note that Tor only aims to protect against traffic analysis attacks such as website fingerprinting [4], [22], it does not aim to protect against traffic confirmation attacks. Nevertheless, we show that highly accurate traffic analysis can be performed cheaply and quantify the amount of anonymity leaked when the full network flow is not seen and under proposed padding defenses [9] that aim to protect against traffic analysis.

## III. Methodology

### A. *Threat Model*

We study communications between a client and a website over an OR network as depicted in Figure 1, where a network flow passes through three relays before reaching its destination. The adversary has the capability to record all network traffic in some local area network, possibly from multiple senders. The adversary also eavesdrops between an exit relay and a website. The ultimate goal is to link the network flow observed after the exit relay to the correct network flow in the LAN observed before reaching the entry relay. All communications in our model are sent simultaneously; if only one network flow was sent and captured during some time period, linking of flows would be trivial. We assume that there is enough diversity in the network that the probability of a relay being an entry and exit for the same network flow is statistically unlikely.

### B. *Data Collection & Feature Selection*

*1) Data Collection:* We use the publicly available Wang et al. data set [22] to test our attacks. The data set consists of 90 instances of 100 sensitive websites that are blocked in countries such as China, UK and Saudi Arabia. This data set was collected via Tor which pads all packets to a fixed size of 512-bytes[1], so the only metadata from which we can leverage information is the time a packet was seen and the direction of the packet.

This data was collected at the client side, to generate network traffic traces that an adversary would capture server side we construct an inter-packet delay variation (IPDV) distribution. The IPDV distribution represents the jitter in the network between observing a website load at the client side and observing a website load at the server side. Modifications are applied to the Wang et al. data set to construct the server side data set as follows:

- **Probability of dropped packet** - 1%, 5%, 10% and 30%.

[1]Commonly referred to as cells.

- **Obfuscation of time stamps** - Each packet timestamp was modified by adding $u$ to it. Where $u$ is randomly drawn from the experimentally derived IPDV distribution .

The IPDV distribution is constructed through real world experiments on the Tor network. We set up a simple webpage hosted on five geo-located `Amazon EC2` instances[2]. For each of the five instances we loaded the webpage 100 times through Tor and recorded the packet time arrivals at both client and sever, giving the inter-packet delay variation that an adversary can expect between client and server when collecting traces over Tor. In line with the threat model, our Tor client was configured to use one exit relay to simulate an adversary collecting traffic at one exit. Guard relays were not used, simulating an adversary collecting traffic over some local network where a client could potentially use any entry node in the network. We found that although the IPDV distribution exhibited a high degree of variance, the average value was close to 0ms and resembled a normal distribution [3].

We decided not to run a live implementation of our attack over Tor as the application of the experimentally derived IPDV distribution to the public data set collected over Tor reliably simulates traffic collected at the server side. If we instead ran and recorded traffic at an exit relay we would either have to (1) configure the exit relay so it accepts no other traffic than our own; this would not accurately capture how the influence of other traffic in the network affects IPDV, or (2) accept other connections and filter them out after traffic has been recorded. Since clients use Tor to preserve their anonymity we decided against capturing background traffic. Our method of simulating network traffic instances at the server side has the best of both worlds, we do not capture other Tor clients traffic but accurately mimic the delay variations produced by loading a website load over Tor.

*2) Feature Selection:* Feature sets are usually constructed based on some prior assumption of importance that may turn out to be false and as a consequence degrade the accuracy of the classifier. Our set of features is designed to be low-level, meaning we make no assumptions about the importance of certain features such as concentrations of incoming or outgoing packets, orderings of packets, and directions of packets.
Once the transmission of a flow has finished, the flow is split in to *N* evenly spaced time windows, where *N* is decided on prior to computation. The number of packets seen in a time window is counted and recorded. This information is then used to create the hash of the network traffic flow. When a network flow contains fewer packets than the total number of time windows, we simply discard this network flow - note that this happened infrequently due to the small time windows chosen.

### C. Coskun-Memon Algorithm

We adapt the algorithm from Coskun et al. [7] to compute the binary hash of a network traffic flow. The algorithm was originally used to identify matching VoIP flows, but has been modified since VoIP flows contain both the time and size of packets.

[2]Instance were hosted in Oregon, Ireland, Tokyo, Sydney, São Paulo.

[3]Inter-packet delay variation minimum = -1418ms, inter-packet delay variation average = 21ms, inter-packet delay variation maximum = 1735ms.

**Data**: network flow, number of time windows $N$
**Result**: binary hash of network flow
$H = [0, 0, ..., 0]$;
Extract number of packets in time windows $T_0$, ..., $T_N$;
**for** $i \leftarrow 0$ **to** $N$ **do**
  **if** $i = 0$ **then**
    $H \leftarrow H$ + $[R_1(\bar{T_0}), ..., R_m(\bar{T_0})]$;
  **end**
  **if** $i > 0$ **then**
    $\bar{B_i} \leftarrow B_i - B_{i-1}$;
    $H \leftarrow H$ + $\bar{B_i}[R_1(\bar{T_i}), ..., R_m(\bar{T_i})]$;
  **end**
  h = *sign*(H)
**end**

**Algorithm 1:** Computes a binary hash of a network traffic flow.

Algorithm 1 describes how we compute the binary hash of a network traffic flow. The algorithm takes a network traffic flow and computes the cumulative number of packets in each time window as described in section III-B.. Then it initializes a list, *H*, of $m$ integers all set at 0. For each time window $T_0$, .., $T_N$ it extracts the time, $\bar{T_i}$, and the number of packets in that window, $B_i$, for $i \in \{0, ..., n\}$. For every time window after the first it computes the relative difference in number of packets seen, $\bar{B_i} = B_i - B_{i-1}$. For each time window it projects $\bar{T_i}$ on to $m$ pseudo-random bases $R_a()$ [4]. The pseudo-random bases were chosen so that packets with similar timings will be projected to values on the bases that are close to one another, resulting in similar hashes for similar network traffic flows. *H* is then updated by multiplying each projection with $\bar{B_i}$. Finally once this has been done for every time window a hash $h$ is produced by setting h = *sign*(H) where:

$$h_a = sign(H_a) = \begin{cases} 1 & \text{if } H_a > 0 \\ 0 & \text{if } H_a \leq 0 \end{cases} \text{ for } a \in \{1, .., m\}.$$

The adversary stores the computed hashes at either end of the network, which will then be used for comparison.

## IV. Attack on Unpadded Mixes

We evaluate the performance of our attack by performing experiments simulating a traffic confirmation attack. First we describe the attack setup and then explain the results of our experiments.

We assume communication is done over an OR network and the adversary can observe traffic at some point between the client and entry relay and again at some point between the exit relay and destination of traffic. Typically an adversary cannot expect to compute the same hash since network impairments such as packet delays will alter the structure of a network flow. Therefore we test the ability for an adversary to successfully link flows under simulated network impairments. Since the

[4]We chose the function

$$R_a(x) = \sin(x + a)/5 + \sin((x + a) \cdot r_a) \cdot r_a$$

where $r_a$ is a unique random value between -1 and 1 and $a \in \{1, ..., m\}$.

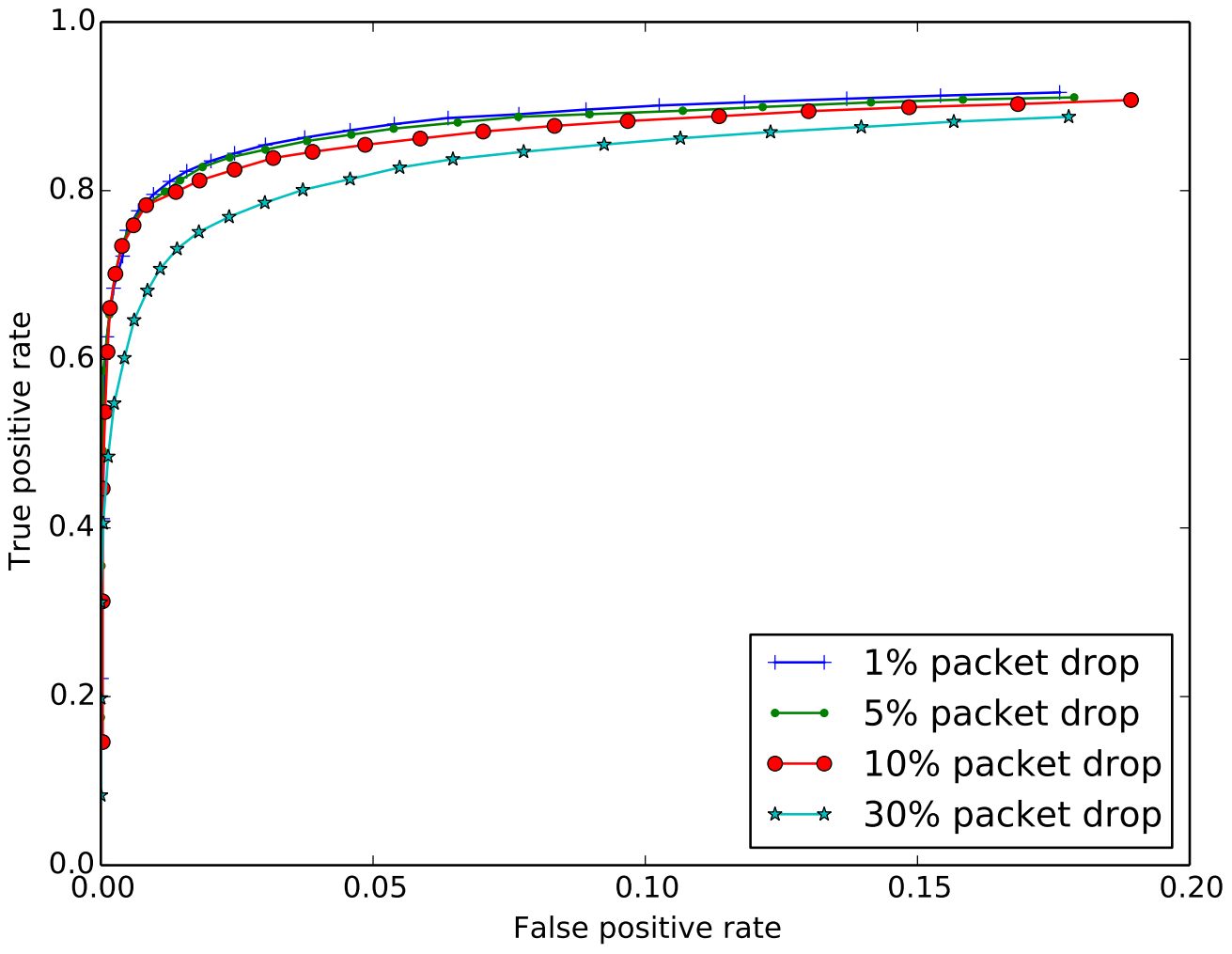

**(a)** ROC for varied fractions of dropped packets. Hash length is set at 256-bits.

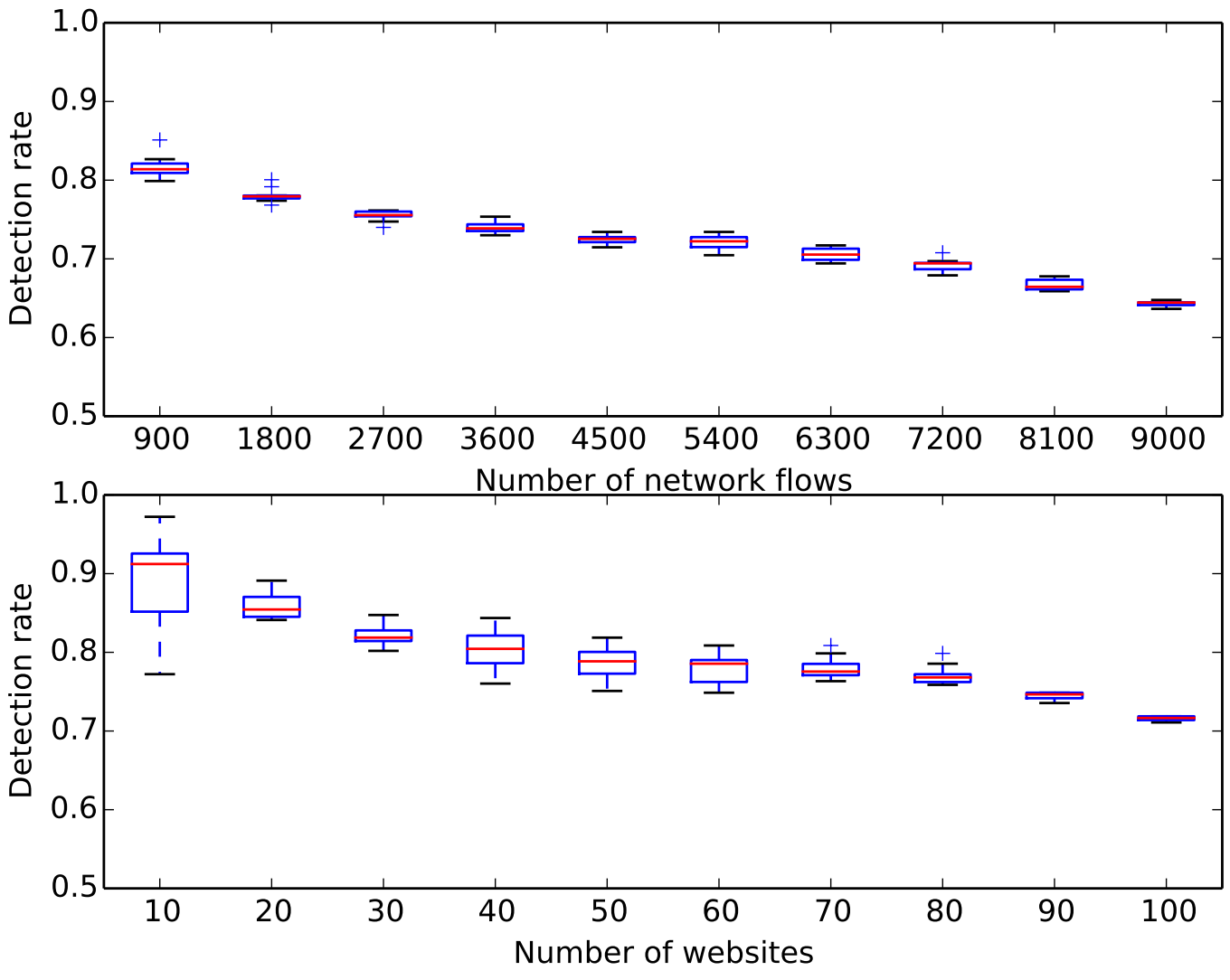

**(b)** The success of perfect network flow matching and website matching for 1% packet drop rate over repeated experiments.

**Fig. 2:** Traffic confirmation attack accuracy.

dataset consists of 90 instances of each of the 100 websites we test the ability of the adapted algorithm to both correctly link a network flow and the ability to correctly link flows from the same website. This is the problem of matching the correct network flow out of 9000 network flows.

We compute the Hamming distance between the hash of the original flow and the hash of the modified flow and mark it a correct identification if the distance is below a threshold. We also compute the Hamming distance between the hash of the original flow and the hash of a random modified flow. If the distance is below the same threshold we mark this a false positive match.

### A. *Results*

Figure 2a gives the ROC curve for different probabilities of dropped packets as the threshold value is changed. Figure 2a shows that match accuracy decreases as the fraction of dropped packets increases. This is to be expected, as the total number of packets shared between the original and modified flows decreases, the Hamming distance between the two hashes will rise resulting in fewer true positive matches. Nevertheless, even if 10% of packets in a network traffic flow are dropped an adversary can correctly match over 80% of network flows with only a 2% false positive rate. We note that Tor uses TCP/IP which guarantees transmission of every packet and so we can expect our attack to be even more successful on Tor than on OR networks using unreliable protocols such as UDP.

Figure 2a shows that an adversary can successfully match hash pairs given some boundary from which to decide if the match is a success. We now consider a stronger metric for classification; for each network flow hash we compute the distance with all modified network flow hashes and check if the hash with minimum distance is its true pair. Our task is made more complicated since our dataset contains many flows that look similar to one another. This is because network flows were generated by loading a small number of websites. We can expect any network flow to look similar to 1%[5] of the entire collection of network flows. Under this new metric of success the following experiment was performed - for each network traffic flow we recorded the IDs of the modified network flows[6] whose hash had minimum Hamming distance. If any of the IDs were the true modified network flow counterpart we considered this a perfect match. If any of the IDs belonged to the same website we considered this a successful match, as an adversary would assign the flow to the correct website. Figure 2b shows the match success rate of linking flows to flows and flows to websites. The rate of attack degradation is gradual as the amount of data grows. If the aim of the attack is to infer which website a client is visiting out of a list of 50 monitored websites (of 90 instances each) then an adversary can expect accurate classification over 80% of the time and drops to 72% when monitoring 100 websites. An adversary can expect perfect network flow matching 72%-74% when monitoring 4500 network flows (50 websites) and drops to 65% when monitoring 9000 network flows (100 websites).

## V. Attack on Padded Mixes

We now consider a traffic confirmation attack over an OR network that uses Adaptive Padding (AP). All flows that are seen before the entry relay are padded by AP with padding removed by some trusted relay in the network before arriving at the destination of the communication. In the opposite direction we assume AP is applied before an adversary captures traffic (either by directly applying AP at the web sever, or at a trusted

[5]90 out of 9000 network traffic flows.

[6]Hash length = 256-bits, probability of dropped packet = 1% and packet timestamp obfuscation added.

bridge) and is removed before an adversary captures traffic at the end of the circuit.

**Data**: one unpadded network flow $U$=$\{U_{in}$,$U_{out}\}$, padded network flows $P_{in}$=$\{P_{in_0}$, ..., $P_{in_l}\}$ and $P_{out}$=$\{P_{out_0}$, ..., $P_{out_l}\}$, length of time window $k$
**Result**: Match prediction between $U$ and $P_t$=$\{P_{in_t}$,$P_{out_t}\}$ for $t \in \{1, ..., l\}$

**for** $i \leftarrow in$ **and** $out$ **do**
  split $U_i$ and $P_i$ in to non-overlapping time windows of size $k$;
  $h$ = [];
  **for** $m \leftarrow 0$ **to** $l$ **do**
    $h_m$ = number of windows with an equal number of packets between $U_i$ and $P_{i_m}$;
    Append $h_m$ to $h$;
  **end**
  $P_{i_x}$ where $x$=*index(max(h))*;
**end**
**if** $x$ is equal for both $P_{in_x}$ and $P_{out_x}$ **then**
  **Return** $U$ = $P_x$
**end**

**Algorithm 2:** How to compute a match between an unpadded flow and padded flows.

### A. Evaluation

We applied AP to network flows at a padding rate of 54% - using AP with a padding rate over 50% was shown by Shmatikov et al. [20] to significantly degrade the performance of a traffic confirmation attack consisting of correlating inter-arrival packet times on two links. For both incoming and outgoing network flows, we created histograms of inter-arrival packet times by crawling the top 25K alexa sites[7]. We then applied AP to each of the 9000 traces in our data set, to simulate the padded flows collected by the adversary at the start of communication. We also apply IPDV in the same fashion as section IV to the 9000 traces to simulate various network impairments in the OR network, these are the simulated unpadded flows the adversary collects. Note that the adversary is only concerned with classifying one unpadded network flow at any one time.

### B. Algorithm

Algorithm 2 presents adapted-SCC, a time window based traffic confirmation attack based on selective cross-correlation (SCC) [2]. Adapted-SCC links an unpadded flow with its padded counterpart. Adapted-SCC is applied by first separating all observed flows in to incoming or outgoing flows - in our case, these both consist of 9000 flows. Since AP inserts dummy packets to gaps in a flow, the padded flow is a super-set of the original flow. Importantly packet times are preserved in both padded and unpadded flows (with some minor packet time variability), allowing for comparison between the number of bursts in the padded flow and the unpadded flow. Similar to section III-B., adapted-SCC splits a flow in to non-overlapping windows of time. The difference is that in section III-B., the number of time windows is predetermined, so all flows are split in to an equal number of sections. Here the length of one time window is predetermined, resulting in different numbers of time windows for different flows. Each padded flow is given a score that corresponds to the number of windows which share an equal number of packets with the unpadded flow. Adapted-SCC takes the score for both incoming and outgoing padded flows, if the highest score of the incoming and outgoing flow is the from the same network traffic flow we consider this padded flow as the match of the unpadded flow.

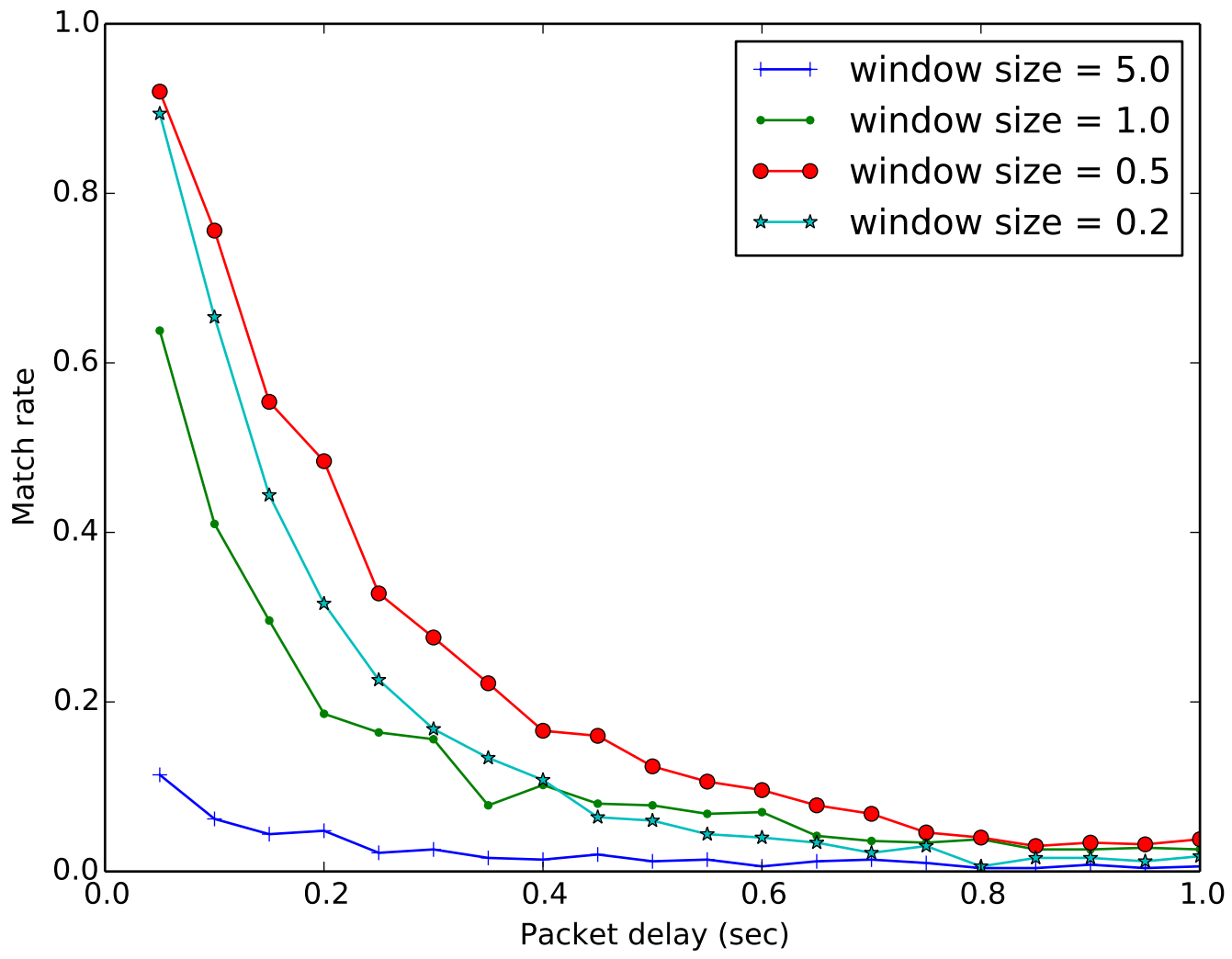

**Fig. 3:** The average successful match rate between all padded flows and an unpadded flow with packet jitter added.

### C. Results

Figure 3 shows the success rate of adapted-SCC against AP as the IPDV average is increased for different window periods. Generally we see that the smaller the window size the more accurate the attack, since a smaller window time is able to capture the bursts in flows that were not padded by AP. With an IPDV average of 200ms an adversary is still able to match over 50% of streams correctly, the match rate increases to over 75% when the IPDV average is 100ms. At the experimentally observed IPDV of 21ms, adapted-SCC is able to match over 90% of flows correctly. We observed almost no false positives since adapted-SCC requires both the padded incoming and outgoing streams to agree to make a prediction. While this did not happen, we observed a rate of failure of prediction almost equal to (1 - match prediction rate).

## VI. Discussion & Conclusion

We consider the threat model where an adversary collects 9000 hashes of network flows simultaneously on the client side, and collects flow hashes over time on the server side, matching against the client side collected data. Tor metrics[8] estimate there to be over two million daily users of Tor and so we estimate that there are enough concurrent connections in Tor for this model to be realistic. Our results show that traffic confirmation attacks are possible and accurate at scale. Using simple packet counting schemes a relatively weak adversary can build robust short fingerprints of network traffic flows to perform powerful traffic analysis attacks when only an incomplete fraction of network traffic is visible or when probabilistic padding schemes are used. We note that packet delays were drawn independently from the IPDV distribution, in reality this will not be the case, packet delays are dependent on the load of the network at the time of request. As a next step a live implementation of the attack could be performed on Tor, but this requires care since we do not wish to capture other clients traffic yet we need to ensure that background traffic is present.

[7]http://www.alexa.com/topsites

[8]https://metrics.torproject.org/userstats-relay-country.html

We note that while our attack on padding schemes faithfully replicates the proposed form of AP applied to an OR network, AP can only be applied to Tor in its current form in the forward direction. Relays do not share session keys with another; session keys are shared between a relay and a client. Relays do not have the ability to generate multi-hop dummy cells, instead clients must provide the dummy cells. The ability for relays to generate dummy cells would also violate Tor's integrity checks since the running digest must be synchronized between client and relays, but a client has no way to check if a relay has generated new dummy cells. Before AP can be applied in Tor either a protocol change must be applied or web servers must be persuaded to generate dummy cells server side, which is clearly an unrealistic request for standard websites but may be tolerable for hidden services.

Probabilistic padding defenses that do not purposefully delay packets are still vulnerable to simple yet powerful timing attacks. We expect to evaluate next alternative padding schemes such as dependent link padding [23] or adding intentional delays to packets. As Figure 3 shows, adapted-SCC becomes ineffective as the volatility of jitter increases. Adding small random packet delays could foil timing attacks without incurring a high latency overhead.

## VII. Acknowledgements

We would like to thank the anonymous reviewers and George Danezis for their helpful comments.

## References

[1] Australian Government Data Retention Policy. https://www.ag.gov.au/NationalSecurity/DataRetention/Documents/Dataset.pdf, 2015. [Online; accessed November-2015].

[2] T. Abraham and M. Wright. Selective cross correlation in passive timing analysis attacks against low-latency mixes. In *Global Telecommunications Conference (GLOBECOM 2010), 2010 IEEE*, pages 1–5, Dec 2010.

[3] Kevin Bauer, Damon McCoy, Dirk Grunwald, Tadayoshi Kohno, and Douglas Sicker. Low-resource routing attacks against tor. In *Proceedings of the 2007 ACM Workshop on Privacy in Electronic Society*, 2007.

[4] Xiang Cai, Xin Cheng Zhang, Brijesh Joshi, and Rob Johnson. Touching from a distance: website fingerprinting attacks and defenses. In *the ACM Conference on Computer and Communications Security, CCS'12, Raleigh, NC, USA, October 16-18, 2012*, pages 605–616, 2012.

[5] Sambuddho Chakravarty, Marco V. Barbera, Georgios Portokalidis, Michalis Polychronakis, and Angelos D. Keromytis. On the effectiveness of traffic analysis against anonymity networks using flow records. In *Proceedings of the 15th International Conference on Passive and Active Measurement - Volume 8362*, 2014.

[6] Sambuddho Chakravarty, Angelos Stavrou, and AngelosD. Keromytis. Traffic analysis against low-latency anonymity networks using available bandwidth estimation. In *Computer Security ESORICS 2010*, pages 249–267. Springer Berlin Heidelberg, 2010.

[7] Baris Coskun and Nasir D. Memon. Tracking encrypted voip calls via robust hashing of network flows. In *Proceedings of the IEEE International Conference on Acoustics, Speech, and Signal Processing, ICASSP 2010, 14-19 March 2010, Sheraton Dallas Hotel, Dallas, Texas, USA*, pages 1818–1821, 2010.

[8] Roger Dingledine, Nick Mathewson, and Paul Syverson. Tor: The second-generation onion router. In *Proceedings of the 13th Conference on USENIX Security Symposium - Volume 13*, SSYM'04, pages 21–21, Berkeley, CA, USA, 2004. USENIX Association.

[9] Nicholas Hopper, Eugene Y. Vasserman, and Eric Chan-Tin. How much anonymity does network latency leak? *ACM Trans. Inf. Syst. Secur.*, 13(2), 2010.

[10] MWR INFOSECURITY. Detecting and deterring data exfiltration. *Guide & Technical Report*, Feb 2014.

[11] Brian Neil Levine, Michael K. Reiter, Chenxi Wang, and Matthew K. Wright. Timing attacks in low-latency mix systems (extended abstract). In *Financial Cryptography, 8th International Conference, FC 2004, Key West, FL, USA, February 9-12, 2004. Revised Papers*, pages 251–265, 2004.

[12] Zhen Ling, Junzhou Luo, Wei Yu, Xinwen Fu, Dong Xuan, and Weijia Jia. A new cell counter based attack against tor. In *Proceedings of the 16th ACM Conference on Computer and Communications Security*, CCS '09, pages 578–589, New York, NY, USA, 2009. ACM.

[13] S. J. Murdoch. Comparison of tor datagram designs. *Tor Project Technical Report*, Nov 2011.

[14] Steven J. Murdoch and George Danezis. Low-cost traffic analysis of tor. In *2005 IEEE Symposium on Security and Privacy (S&P 2005), 8-11 May 2005, Oakland, CA, USA*, pages 183–195, 2005.

[15] Steven J. Murdoch and Piotr Zielinski. Sampled traffic analysis by internet-exchange-level adversaries. In *Privacy Enhancing Technologies, 7th International Symposium, PET 2007 Ottawa, Canada, June 20-22, 2007, Revised Selected Papers*, pages 167–183, 2007.

[16] Pai Peng, Peng Ning, and Douglas S. Reeves. On the secrecy of timing-based active watermarking trace-back techniques. In *IEEE Symposium on Security and Privacy*, pages 334–349. IEEE Computer Society, 2006.

[17] Joel Reardon and Ian Goldberg. Improving tor using a tcp-over-dtls tunnel. In *Proceedings of the 18th Conference on USENIX Security Symposium*, SSYM'09, pages 119–134, Berkeley, CA, USA, 2009. USENIX Association.

[18] Intel Security. Grand theft data. data exfiltration study: Actors, tactics, and detection. *Technical Report*, Sep 2015.

[19] Andrei Serjantov and Peter Sewell. Passive attack analysis for connection-based anonymity systems. In *In Proceedings of European Symposium on Research in Computer Security (ESORICS*, pages 116–131, 2003.

[20] Vitaly Shmatikov and Ming-Hsiu Wang. Timing analysis in low-latency mix networks: Attacks and defenses. In *ESORICS*, Lecture Notes in Computer Science. Springer, 2006.

[21] Yixin Sun, Anne Edmundson, Laurent Vanbever, Oscar Li, Jennifer Rexford, Mung Chiang, and Prateek Mittal. RAPTOR: routing attacks on privacy in tor. *CoRR*, abs/1503.03940, 2015.

[22] Tao Wang, Xiang Cai, Rishab Nithyanand, Rob Johnson, and Ian Goldberg. Effective attacks and provable defenses for website fingerprinting. In *Proceedings of the 23rd USENIX Security Symposium, San Diego, CA, USA, August 20-22, 2014.*, pages 143–157, 2014.

[23] Wei Wang, Mehul Motani, and Vikram Srinivasan. Dependent link padding algorithms for low latency anonymity systems. In *Proceedings of the 15th ACM Conference on Computer and Communications Security*, CCS '08, pages 323–332, New York, NY, USA, 2008. ACM.

[24] X.Fu and Z.Ling. One cell is enough to break tors anonymity. In *Proceedings of Black Hat Technical Security Conference,*, 2009.